\documentclass[12pt]{article}
\usepackage{amsmath,amssymb,bm,amsthm}
\usepackage{graphicx,xcolor}
\usepackage{array,cite,url}
\usepackage{amsfonts,epstopdf}
\usepackage{booktabs}
\usepackage{array}
\usepackage[shortlabels]{enumitem}
\usepackage[bookmarks=false]{hyperref} 
\usepackage{cleveref}
\usepackage{indentfirst}
\newcommand{\p}{\partial}
\newcommand{\x}{\mathbf{x}}

\newcommand{\rar}{\rightarrow}
\newcommand{\bs}{\boldsymbol}
\newcommand{\lf}{\text{lift}}

\newcommand{\Kc}{\mathcal{K}}
\newcommand{\mr}[1]{\mathrm{#1}}

\usepackage{algorithm}
\usepackage[noend]{algpseudocode}
\makeatletter
\def\BState{\State\hskip-\ALG@thistlm}
\makeatother
\usepackage{algorithmicx}

\title{\LARGE \bf
	A data-driven Koopman model predictive control framework for nonlinear flows}

\author{Hassan Arbabi, Milan Korda and Igor Mezi\'c
	\thanks{The authors are with the department of Mechanical Engineering,
		University of California, Santa Barbara, CA, 93106, USA. {\tt\small\{harbabi, milan.korda, mezic\}@engineering.ucsb.edu}}%
}

\hypersetup{
  colorlinks   = true, 
  urlcolor     = blue, 
  linkcolor    = red, 
  citecolor   = blue 
}

\newif\ifcomments
\newcommand\comments[1]{\ifcomments\textcolor{blue}{#1}\else\relax\fi}
\commentsfalse

\begin{document}

\maketitle

\begin{abstract}
The Koopman operator theory is an increasingly popular formalism of dynamical systems theory which enables analysis and prediction of the nonlinear dynamics from measurement data. Building on the recent development of the Koopman model predictive control framework~\cite{korda2018linear}, we propose a methodology for closed-loop feedback control of nonlinear flows in a fully data-driven and model-free manner.  In the first step, we compute a Koopman-linear representation of the control system using a variation of the extended dynamic mode decomposition algorithm and then we apply model predictive control to the constructed linear model. Our methodology handles both full-state and sparse measurement; in the latter case, it incorporates the delay-embedding of the available data into the identification and control processes. We illustrate the application of this methodology on the periodic Burgers' equation and the boundary control of a  cavity flow governed by the two-dimensional incompressible Navier-Stokes equations\footnote{The MATLAB implementation of the Koopman-MPC framework for the examples is available at \url{https://github.com/arbabiha/KoopmanMPC_for_flowcontrol}. }. In both examples the proposed methodology is successful in accomplishing the control tasks with sub-millisecond computation time required for evaluation of the control input in closed-loop, thereby allowing for a real-time deployment.
\end{abstract}



\section{Introduction}
Flow control is one of the central topics in fluid mechanics with an enormous impact on other fields of engineering and applied science. Its wide range of applications includes, just to name a few, reduction of aerodynamic drag on vehicles and aircrafts, mixing enhancement in combustion and chemical processes, suppression of instabilities to avoid structural fatigue, lift increase for wind turbines, and design of biomedical devices. To emphasize the impact of flow control, it is worth noting that discovery of efficient flow control techniques for reduction of  drag on ships and cars can result in mitigation of yearly CO$_2$ emission by millions of tons and annual savings of billions of dollars in the global shipping industry \cite{kim2011physics,brunton2015closed}.

Despite all the interest and continuous effort, the flow control still poses a daunting challenge to our theoretical understanding and computational resources. The main source of difficulty is the combination of high-dimensionality and nonlinearity of fluid phenomena which results in computational or experimental models which are too complex and costly to control using well-developed strategies of modern control theory. The recent advances in numerical computation has led to partial success with active control of flows using models based on Navier-Stokes equations \cite{bewley2001flow,bewley2001dns,kim2007linear,kim2011physics}; however, these methods suffer from two major shortcomings: first, nonlinear models obtained from Navier-Stokes are still high-dimensional and computationally costly, thereby not allowing for fast implementation of nonlinear and computationally complex control techniques such as nonlinear model predictive control, and second, linear models used with LQR/LQG or adjoint-based controllers often rely on local linearization around equilibria or a trajectory of the flow which makes them valid only locally, and may result in suboptimal or even unstable control performance.

\comments{Hassan: I have made some changes to the above paragraph}

An alternative approach that has gained traction in the last two decades is identification of relatively low-dimensional flow models from data provided by numerical simulations or experiments. Some of the data-driven methods combine the measurement data with underlying physical model to identify models of the system. The major examples include construction of (usually autonomous) state space models via Galerkin projection of the Navier-Stokes equations onto the modes obtained by proper orthogonal decomposition (POD) of data \cite{holmes2012turbulence,noack2003hierarchy,balajewicz2013low},  or identification of linear input-output systems using balanced POD \cite{willcox2002balanced,rowley2005model}. There are also a few applications of system identification methods to construct linear input-output models purely from data, including the eigensystem realization algorithm \cite{cabell2006experimental,brunton2013reduced}, as well as subspace identification and autoregressive models \cite{huang2008control,herve2012physics}.
Utilization of the above techniques in a variety of problems has shown great promise for low-dimensional modeling  and control of complex flows from data.

In this paper, we present a general and fully data-driven framework for control of nonlinear flows based on the Koopman operator theory \cite{koopman1931,mezic2013analysis}. This theory is an operator-theoretic formalism of classical dynamical systems theory with two key features: first, it allows a scalable reconstruction of the underlying dynamical system from measurement data, and second, the models obtained are linear (but possibly high-dimensional) due to the fact that the Koopman operator is a linear operator whether the dynamical system is linear or not. The linearity of the Koopman models is especially advantageous since it makes them amenable to the plethora of mature control strategies developed for linear systems. This framework for design of controller, is called  \emph{Koopman-MPC} and follows the work in \cite{korda2018linear}. In the first step of our approach, we build a finite-dimensional approximation of the controlled Koopman operator from the data, using a variation of the extended dynamic mode decomposition algorithm (EDMD) \cite{williams2015data}, with a particular choice of observables assuring linearity of the resulting approximation. The ideal data would include measurements on a number of system trajectories with various input sequences.  In the second step, we apply the model predictive control (MPC) to these linear models to obtain the desired objectives. The distinguishing feature is that this framework leads to a \emph{linear} MPC, solving a convex quadratic programming problem, and thereby enables a rapid solution of the underlying optimization problem, which is necessary for real-time deployment. This methodology can be also applied to problems with sparse measurements, i.e., problems with a limited number of instantaneous measurements (e.g. point measurements of the velocity field at several different locations).  In that case, delay-embedding of the available measurements (and nonlinear functions thereof) is used to construct the Koopman-linear model; the MPC is then applied to the linear system whose state variable is  the delay-embedded vector of measurements.

 The outline of this paper is as follows: A brief review of related work is given in  \cref{sec:lit}. \Cref{sec:KoopmanTheory} gives a review of the Koopman operator theory for dynamical systems with input. In \cref{sec:EDMD}, we describe the EDMD algorithm for construction of the Koopman-linear model from measurement data. In \cref{sec:DelayEmbed}, we discuss using delay-embedding to construct and control Koopman-linear models from sparse measurements. An overview of the MPC framework is given in \cref{sec:MPC}.
In \cref{sec:Examples}, we present two numerical examples: the Burgers' system on a periodic domain and the 2D lid-driven cavity flow. We formulate the control problem for these cases using various objectives and demonstrate our approach for both full-state and sparse measurements. We summarize the results and discuss the outlook in \cref{sec:conclusion}.

\subsection{Review of related work}\label{sec:lit}

The Koopman operator formalism of dynamical systems is rooted in the seminal works of Koopman and Von Neumann in the early 1930s \cite{koopman1931,koopmanandvonneumann:1932}. This formalism  appeared mostly in the context of ergodic theory for much of the last century, until in mid 2000's when the works in \cite{mezic2004comparison,mezic2005} pointed out  its potential for rigorous analysis of dynamical systems from data. The notion of \emph{Koopman mode decomposition (KMD)} which is based on the expansion of observable fields  in terms of Koopman operator eigenvalues and eigenfunctions was also introduced in \cite{mezic2005}.
KMD was first applied to a complex flow in \cite{rowley2009} where its connection with the DMD numerical algorithm \cite{schmid2010} was  pointed out. The work in \cite{rowley2009} showed the promise of this viewpoint in  extracting the physically relevant flow structures and time-scales from data.  Following the success of this work, KMD and its numerical implementation through DMD, has become a popular decomposition for dynamic analysis of nonlinear flows \cite{schmid2011applications,hua2016dynamic,bagheri2013koopman,sayadi2014reduced,subbareddy2014direct}.

The application of the Koopman operator to data-driven control of high dimensional systems is much less developed. The earliest works on generalizing the Koopman operator approach to control systems was presented in \cite{Proctor2016generalize,koopman_cont_extend} accompanied with a numerical variation of DMD algorithm \cite{proctor2016dynamic}. To the best of our knowledge, however, the only application for feedback control of fluid flows are the works in \cite{peitz2017koopman,peitz2018controlling}.
The work in \cite{peitz2017koopman} considered the problem of flow control using a finite set of input values. For each value of the input, a Koopman-linear model was constructed from the data and the control problem was formulated as a switched optimal control and implemented in a receding horizon fashion. This methodology was successfully used for tracking reference output signals in Burgers equation and incompressible flow past a cylinder. The work in \cite{peitz2018controlling} proposed to remove the restriction of the input to a finite set, by  interpolating between the Koopman-linear systems for each  input value which led to an improvement of the control performance.
In \cite{korda2018linear} (which this work is based on), a more general extension of the Koopman operator theory to systems with input was presented, and used to construct linear predictors especially suitable for model predictive control, demonstrating the effectiveness of the approach (among other examples) on the control of the Korteweg-de Vries PDE. The results in \cite{korda2018linear} showed superiority of the controlled Koopman-linear predictors constructed from data to models obtained by local linearization and Carleman's representation both for prediction and for feedback control. Let us also mention the earlier work in \cite{Glaz2017quasi} that utilized KMD to construct the normal forms, for dynamics of the flow past an oscillating cylinder; the input forcing appeared as a bilinear term in the normal forms for this flow.  See \cite{sootla2017optimal} for application to pulse-based control of monotone systems as well as  \cite{mauroy2017koopman} for system identification  and \cite{surana_estim} for state estimation.

The numerical engine behind the system identification part of the framework presented in this paper is the Extended Dynamic Mode Decomposition (EDMD) algorithm proposed in \cite{williams2015data}.
Although the original DMD algorithm was invented independent of the Koopman operator theory \cite{schmid2010}, the connection between the two was known from early on \cite{rowley2009}, and DMD-type algorithms have become the popular methods for computation of the Koopman operator spectral properties. Nevertheless,
the convergence of DMD algorithms for approximation of Koopman operator is just recently established in \cite{arbabi2017ergodic,korda2017convergence}.
For application of Koopman-MPC to systems with sparse measurements, the EDMD is modified to include the delay embeddings of instantaneous measurements.
 Delay embedding is a classic technique in system identification literature (see, e.g., ~\cite{ljung1998system} for a comprehensive reference) and control as well as in linear and nonlinear time-series analysis (e.g., \cite{tjostheim1994nonparametric}). In the field of dynamical systems, the classical reference is the work of Takens~\cite{takens1981detecting} on geometric reconstruction of nonlinear attractors. The work in \cite{tu2014dynamic} suggested the combination of this technique with the DMD algorithm for identification of nonlinear systems and its role in approximation of the Koopman operator was studied in \cite{arbabi2017ergodic,korda2017data}; the use for control, in the Koopman operator context, was described in~\cite{korda2018linear}.

\section{Koopman operator theory}\label{sec:KoopmanTheory}
In this section, we first review the basics of the Koopman operator formalism for autonomous dynamical systems and then discuss its extension to systems with input and output. We will  focus on \textit{discrete-time} dynamical systems to be consistent with the discrete-time nature of the measurement data, but most of the analysis easily carries over to the continuous-time systems. We refer the reader to \cite{mezic2013analysis,budisic2012applied} for a more detailed discussion of the Koopman operator basics.

Consider the dynamical system
\begin{align}
x^+=T(x),\quad x\in M
\end{align}
defined on a state space $M$. We call any function $g:M\rar \mathbb{R}$ an \textit{observable} of the system, and we note that the set of all observables forms a (typically infinite-dimensional) vector space. The Koopman operator, denoted by $\Kc$, is a linear transformation on this vector space given by
\begin{align}\label{eq:KoopmanDef}
\Kc g = g\circ T,
\end{align}
where $\circ$ denotes the function composition, i.e., $(\Kc g)(x)=g(T(x))$. Informally speaking, the Koopman operator updates the observable $g$ based on the evolution of the trajectories in the state space. The key property of the Koopman operator that we exploit in this work is its linearity, that is, for any two observables $g$ and $h$, and scalar values $\alpha$ and $\beta$,  we have
\begin{align}
\Kc(\alpha g + \beta h)=\alpha \Kc  g + \beta \Kc h,
\end{align}
which follows from the definition in \cref{eq:KoopmanDef}. We call the observable $\phi$ a Koopman eigenfunction associated with Koopman eigenvalue $\lambda \in \mathbb{C}$ if it satisfies
\begin{align}
\Kc\phi=\lambda \phi.
\end{align}
The spectral properties of the Koopman operator can be used to characterize the state space dynamics; for example, the Koopman eigenvalues determine the stability of the system and the level sets of certain Koopman eigenfunctions carve out the invariant manifolds and isochrons \cite{mauroy2016global,Mezic:2015,mauroy2012use}. Moreover, for smooth dynamical systems with simple nonlinear dynamics, e.g., systems that possess hyperbolic fixed points, limit cycles and tori, the evolution of observables can be described as a linear expansion in Koopman eigenfunctions \cite{mezic2017koopman}. In these systems, the spectrum of the Koopman operator consists of only point spectrum (i.e. eigenvalues) which fully describes the evolution of observables;
\begin{align}\label{eq:KoopmanExpansion}
\Kc^ng=\sum_{j=0}^{\infty}v_j\phi_j \lambda_j^n.
\end{align}
where $v_j$ is called the Koopman mode associated with Koopman eigenvalue-eigenfunction pair $(\lambda_j,\phi_j)$ and it is given by the projection of  the observable $g$ onto $\phi_j$. See \cite{rowley2009} for more detail on Koopman modes, and \cite{mezic2017koopman} on the expansion in (\ref{eq:KoopmanExpansion}).

The extension of the Koopman operator theory to a controlled system denoted by
\begin{align}\label{eq:sysCont}
x^+=T(x,u),\quad x\in M,~u\in \mathcal{U},
\end{align}
 requires one to work on the \textit{extended state space}, which is the Cartesian product of the state space $M$ and the space of all input sequences $\ell(\mathcal{U}) = \{(u_0)_{i=0}^\infty \mid u_i \in \mathcal{U}\}$.
 We denote the extended state space  by $S=M\times \ell(\mathcal{U})$.
 Now, given an observable $g:S\rar\mathbb{R}$ we can define the non-autonomous Koopman operator,
\begin{align}\label{eq:KoopmanDefInput}
(\Kc g)(x,(u_i)_{i=0}^\infty)=g(T(x,u_0),(u_i)_{i=1}^\infty).
\end{align}
See~\cite{korda2018linear} for more details on this extension. We emphasize that the linear representation of the nonlinear system by the Koopman operator is globally valid and generalizes the local linearization around equilibria \cite{mezic2017koopman}.


\section{Construction of Koopman-linear system} \label{sec:EDMD}
In this section, we review the construction of the Koopman-linear system as proposed by \cite{korda2018linear} using the EDMD algorithm \cite{williams2015data}.
We are looking to approximate the dynamics of the nonlinear flow via a linear time-invariant system such as
\begin{align}\label{eq:predictor}
  z^+&=Az+Bu\qquad z\in \mathbb{R}^{n},~u\in \mathbb{R}^k, \notag
  \\ \hat{x} &=Cz.
\end{align}

Consider the set of states and inputs of the nonlinear system in the form of
\begin{equation}\label{eq:StateInput}
  X=[x_1, \ldots, x_K],\quad X^+=[x_1^+~,x_2^+,\ldots,x_K^+],
    \qquad U=[u_1,\ldots,u_K].
\end{equation}
where $x_j^+=T(x_j,u_j)$. Let
\begin{align}
\bs{g}(x)=
\left[
\begin{matrix}
g_1(x)& \ldots & g_{m}(x)
\end{matrix}
\right]^\top
\end{align}
be a given vector of possibly nonlinear observables. These functions may represent user-specified nonlinear functions of the state as well as physical measurements (i.e., outputs) taken on the dynamical system (or nonlinear functions of such outputs).
We are going to assume that we only have access to values of the observables, and therefore, explicit knowledge of the state variable in \eqref{eq:StateInput} is not required.
By collecting data on the dynamical system,  we can form the lifted snapshot data matrices
\begin{align}\label{eq:snapshotmatrices}
  X_\lf&=[\bs{g}(x_1), \ldots, \bs{g}(x_K)],\quad X^+_\lf=[\bs{g}(x^+_1), \ldots, \bs{g}(x^+_K)], \qquad U=[u_1,\ldots,u_K].
\end{align}
These data matrices are the lifted coordinates of the system in the space of observables. Note that, as in~\cite{korda2018linear}, we have not lifted $U$ coordinates to preserve the linear dependence of the predictor on the original input.  The matrices $A$, $B$ and $C$ are then given by the solution to the linear least-squares problems
\begin{align}\label{eq:Min}
  \min_{A,B}\|X^+_\lf-AX_\lf-BU\|_F,\quad   \min_{C}\|X-CX_\lf\|_F
\end{align}
where $\|\cdot\|_F$ denotes the Frobenius norm. The analytical solution to these two problems can be compactly written as
 \begin{align}
  \begin{bmatrix}A & B \\ C & 0 \end{bmatrix}= \begin{bmatrix}X^+_{\lf} \\ X\end{bmatrix}
  \begin{bmatrix}
    X_\lf \\
    U
  \end{bmatrix}^\dagger.
\end{align}
When snapshot matrix $X_\lf$ is fat (i.e. number of columns exceeds number of rows), it is more efficient to compute the matrices by solving the normal equations
 \begin{align}\label{eq:normEq}
  V= \mathcal{M}G,
\end{align}
with the unknown matrix variable $\mathcal{M}$ and given matrices
 \begin{align*}
  V=\begin{bmatrix}X^+_{\lf} \\ X\end{bmatrix} \begin{bmatrix}
    X_\lf \\
    U
  \end{bmatrix}^\top
 ,\quad
  G=\begin{bmatrix}
    X_\lf \\
    U
  \end{bmatrix}
  \begin{bmatrix}
    X_\lf \\
    U
  \end{bmatrix}^\top.
\end{align*}
The solution $\mathcal{M}$ to~(\ref{eq:normEq}) provides the matrices $A$, $B$, $C$ through
\[
\mathcal{M} = \begin{bmatrix}A & B \\ C & 0 \end{bmatrix}.
\]
The matrices  $A$ and $B$ describe the linear dynamics of the Koopman-linear state $z=\bs g(x)$. The prediction of the original state $x$ is obtained simply by $\hat{x} = Cz$.  See \cite{korda2017convergence}  for a convergence analysis of EDMD for approximation of Koopman operator.

\subsection{Sparse measurements and delay embedding} \label{sec:DelayEmbed}
When the number of observables measured on a dynamical system is insufficient for construction of an accurate model, we can use the delay embedding of the observables. Delay
embedding (i.e., embedding several consecutive output measurements into a single data point) is a classical technique ubiquitous in system identification literature (e.g.,~\cite{ljung1998system}) but also in the theory of dynamical systems for geometric reconstruction of nonlinear attractors \cite{takens1981detecting}. It has also been utilized in the context of Koopman framework in \cite{mezic2004comparison,giannakis2017data,arbabi2017ergodic}. The key feature of delay embedding here is that it provides samplings of extra observables to realize the Koopman operator. To be more precise, if we have a sequence of  measurements  on a single observable $h$ at the $n_d$ time instants $t_i,~t_{i+1},\ldots,t_{i+n_d-1}$, we can think of them as sampling of the $n_d$ observables $[h,~\Kc h,\ldots,~\Kc ^{n_d-1}h]$ at the single time instant $t_i$. Here, we describe how we can incorporate delay-embedding into identification and control of Koopman-linear models.
The only requirement for identification is that we should have access to at least $n_d+1$ \emph{sequential} time samples on the trajectories where $n_d$ is the chosen number of delays.

Let $\bs{h} $ be the vector of instantaneously measured observables on the dynamical system (e.g., point measurements of the velocity field), and $n_d$ be the delay embedding dimension.  Consider the state and input matrices described in \eqref{eq:StateInput}, but now assume that they contain a string of sequential samples with length $n_d+1$, i.e., for some $j$, we have
\begin{align}
  x_{i+1}=T(x_i,u_i),\quad i=j,\ldots,j+n_d-1.
\end{align}
We can delay embed the measurements on this string to construct a pair of lifted coordinates in the space of observables,
\begin{align}\label{eq:temp2}
\zeta_j =
\begin{bmatrix}
\bs{h}(x_i)
\\ \vdots \\ \bs{h}(x_{i+n_d-1})
\\ u_i \\ \vdots\\ u_{i+n_d-1}
\end{bmatrix},\quad
\zeta_j^+ =
\begin{bmatrix}
\bs{h}(x_{i+1})
\\ \vdots \\ \bs{h}(x_{i+n_d})
\\ u_{i+1} \\ \vdots\\ u_{i+n_d}
\end{bmatrix},\quad i=j,\ldots,j+n_d-1.
\end{align}

It is easy to check that $\zeta^+_j=\Kc \zeta_j$.
By delay-embedding the observations on all sequential strings of data, we can form the new matrices
\begin{align}\label{eq:temp3}
\tilde{X}=[\zeta_1,~\zeta_2,\ldots,~\zeta_L],\quad \tilde{Y}=[\zeta_1^+,~\zeta_2^+,\ldots,~\zeta_L^+].
\end{align}
We can once again lift the data using a vector of nonlinear user-specified functions $ {\bs {g}}$ to form the new lifted matrices,
\begin{align}\label{eq:temp4}
{X_\lf}&=[ {\bs {g}}(\zeta_1),~ {\bs {g}}(\zeta_2),\ldots,~ {\bs {g}}(\zeta_L)], \notag
\\ \quad Y_\lf&=[ {\bs {g}}(\zeta_1^+),~ {\bs {g}}(\zeta_2^+),\ldots,~ {\bs {g}}(\zeta_L^+)].
\end{align}

Having $X_\lf,~Y_\lf$ and input matrix $U$ defined, we solve the the least-squares problems~(\ref{eq:Min})  to find the linear system matrices. It is very important for the lifting function $ {\bs g}$ to have a meaningful dependence on $u_i,\ldots,u_{i+n_d}$. This allows EDMD to approximate the dynamics of the \emph{extended} state space and discern the effect of previous inputs in the evolution of the state.

The linear predictor in this case would be
\begin{align}\label{eq:predictorEmbed}
  z^+&=Az+Bu \notag
  \\ \hat{\zeta} &=Cz,
\end{align}
Here $\hat\zeta$ denotes the prediction of the ``embedded'' state $\zeta$ (note that when  $\hat\zeta$ is employed for controller design, typically only the part of $\hat \zeta$ corresponding to the most recent output prediction is used).


\section{Model predictive control}\label{sec:MPC}
The methodology presented in the last section allows us to construct a model of the flow in the form of a linear dynamical system (\ref{eq:predictor}).
In this work, we will apply MPC to this linear model to control the original nonlinear flow, but other techniques from modern control theory could be applied as well; see the survey~\cite{mayne2000constrained} or the book \cite{grune2011nonlinear} for an overview of MPC. In the context of MPC, we formulate the control objective as minimization of a cost function over a finite-time horizon. The general strategy is to use the model in \cref{eq:predictor} to predict the system evolution over the horizon, and use these predictions to compute the optimal input sequence minimizing the given cost function along this horizon. Then we \emph{apply only the first element} of the computed input sequence to the real system, thereby producing a new value of the output, and repeat the whole process. This technique is sometimes called the \emph{receding horizon control}. In the following, we describe the notation and some mathematical aspects of this technique. The distinguishing feature when using the lifted linear predictor (\ref{eq:predictor}) is that the resulting MPC problem is a convex quadratic program (QP) despite the original dynamics being nonlinear. In addition, the complexity of solving the quadratic problem can be shown to be independent of the size of the lift if the so-called \emph{dense form} is used \cite{korda2018linear}, thereby allowing for a rapid solution using highly efficient QP solvers tailored for linear MPC applications (in our case qpOASES~\cite{ferreau2014qpoases}).

Let $N$ be the length of the prediction horizon, and $\{u_i\}_{i=0}^{N-1}$ and $\{y_i\}_{i=1}^{N}$ denote the sequence of input and output values over that horizon. A very common choice of cost functions is the convex quadratic form,
\begin{align}\label{eq:MPCcost}
J\big(\{u_i\}_{i=0}^{N-1},\{y_i\}_{i=1}^{N}\big) = &~~y_{N}^{\top}Q_Ny_N +q^{\top} y_N \\
&+ \sum_{i=1}^{N-1} y_i^{\top}Q_i y_i + u_i^{\top}R_iu_i+q_i^{\top} y_i+r_i^{\top}u_i \notag\\
&+ u_0^{\top}R_0u_0+r_0^{\top}u_0, \notag
\end{align}
where $Q_{i=0,\ldots,N}$ and $R_{i=0,\ldots,N-1}$ are real symmetric positive-definite matrices.
The above cost function can be used to formulate many of the common control objectives including the tracking of a reference signal. For example, assume that we want to control the flow such that its output measurements follow an arbitrary time-dependent output sequence denoted by $\{y^{\mr{ref}}\}_i$.
We can formulate this objective as minimization of the distance between $\{y\}_i$ and $\{y^{\mr{ref}}\}_i$, and the corresponding cost function over the finite horizon would be
\begin{align}\label{eq:MPCcost1}
J_1\big(\{u_i\}_{i=0}^{N-1},\{y_i\}_{i=1}^{N}\big) &= \sum_{i=1}^{N} \big(y_i-y^{\mr{ref}}_i\big)^{\top}Q\big(y_i-y^{\mr{ref}}_i\big),\\
& =   \sum_{i=1}^{N} y_i^{\top} Q y_i   - 2\big(y^{\mr{ref}}_i\big)^{\top}Q y_i +\big(y^{\mr{ref}}_i\big)^{\top}Q y^{\mr{ref}}_i.  \notag
\end{align}
where $Q$ is the weight matrix that determines the relative importance of measurements in $y$. Note that the last term in the above equation is not dependent on the input or output, and therefore it does not affect the optimal solution. By dropping this term, and letting $q_i= - Q^{\top}y^{\mr{ref}}_i$, we obtain
\begin{align}\label{eq:MPCcost2}
J_1\big(\{u_i\}_{i=0}^{N-1},\{y_i\}_{i=1}^{N}\big)
& =   \sum_{i=1}^{N} y_i^{\top} Q y_i  +q_i^{\top}y_i,
\end{align}
which is a special form of \cref{eq:MPCcost}. In the numerical examples presented in this paper, we will use this type of cost function.

The MPC controller solves the following optimization problem at \emph{each time step} of the closed loop operation
\begin{align}\label{eq:MPC}
\big(\{u_i^\star\}_{i=0}^{N-1},\{y_i^\star\}_{i=1}^{N}\big)&=\arg\min J\big(\{u_i\}_{i=0}^{N-1},\{y_i\}_{i=1}^{N}\big) \nonumber \\
\text{ {s.t.}} & \qquad  \quad z_{i+1}=Az_i+Bu_i,\quad i=0,\ldots,N \nonumber \\
& \qquad \quad y_i = Cz_i \\
& \qquad \quad E_i^y y_i + E_i^u u_i  \leq b_i,~i=0,\ldots,N-1, \nonumber \\
& \qquad \quad  E_N y_N \leq b_N \nonumber  \\
& \qquad \quad  z_0 = {  {\bs g}}(\zeta_c) \nonumber,
\end{align}
where $\zeta_c$ is the delay-embedded vector of measurements
\begin{align}\label{eq:temp5}
\zeta_c =
[
\bs{h}(x_{k-n_d+1}),\ldots, \bs{h}(x_{k}), u_{k-n_d} , \ldots, u_{k-1}
]^\top
\end{align}

The matrices $E_{i=0,\ldots,N-1}^x$, $E_{i=0,\ldots,N-1}^u$ and $E_N$ define polyhedral state and input constraints. This is a standard form of a convex quadratic programming problem which can be efficiently solved using many available QP solvers - in our case qpOASES~\cite{ferreau2014qpoases}. The computational complexity can be further reduced by expressing the lifted state variables $z$ in terms of the control inputs $u$, thereby eliminating the dependence on the possible very large dimension of $z$; see~\cite{korda2018linear} for details.

\begin{figure}[t!]
\begin{picture}(300,230)
 \put(0,0){\centerline{\includegraphics[width=.9\textwidth]{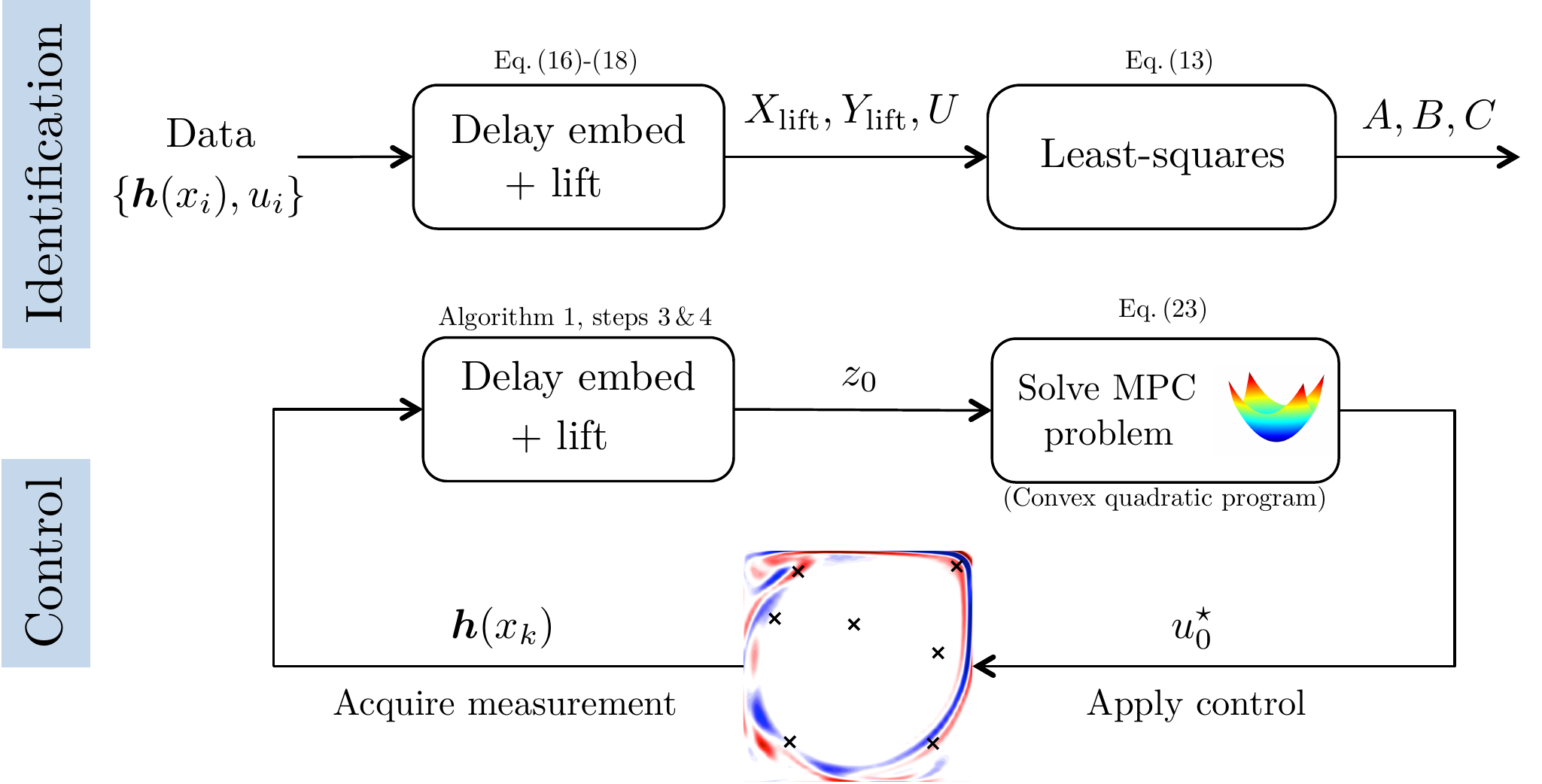}}}
\end{picture}
	\caption{\footnotesize \textbf{Schematic representation of Koopman-MPC framework for  identification and closed-loop control of nonlinear flows.}}
	\label{fig:BigPic}
\end{figure}

Once the optimal input sequence $\{u_i^\star\}_{i=0}^{N-1}$ is computed, we apply its first element $u_0^\star$ to the system to obtain a new output measurement which updates the current state $\zeta_c$; the whole process is then repeated in a receding horizon fashion. \Cref{algo:MPC} summarizes the closed-loop control operation, and the entire algorithm for implementation of the Koopman-MPC is illustrated in \Cref{fig:BigPic} .

\begin{algorithm}
\caption{Koopman MPC -- closed-loop operation}\label{algo:MPC}
\begin{algorithmic}[1]
\Require $\bs{h}(x_{-n_d}),\ldots,\bs{h}(x_{-1}), u_{-n_d},\ldots,u_{-1}$
\For{$k=0,1,\ldots$}
\State Measure $\bs{h}(x_{k})$.
\State Set $\zeta_c = [
\bs{h}(x_{k-n_d+1}),\ldots, \bs{h}(x_{k}), u_{k-n_d} , \ldots, u_{k-1}
]^\top$.
\State Set $z_0 :=  {\bs g}(\zeta_c)$
\State		Solve~(\ref{eq:MPC}) to get an optimal solution $(u_i^\star)_{i=1}^N$
\State Apply $u_1^\star$ to the nonlinear system
\EndFor
\end{algorithmic}
\end{algorithm}

\section{Numerical Examples
\protect\footnote{The MATLAB implementation of the examples is available at \protect\url{https://github.com/arbabiha/KoopmanMPC_for_flowcontrol}.}
}\label{sec:Examples}
\subsection{Burgers equation}

As the first example, we consider the Burgers equation with periodic boundary condition,
\begin{align}\label{eq:Burgers}
 \frac{\p v}{\p t} + v\frac{\p v}{\p z}&= \nu \frac{\p^2 v}{\p z^2}+ f(z,t),\quad z\in[0,1],~t\in[0,\infty). \\
 v(0,t)&=v(1,t)
\end{align}
where $\nu$ is the kinematic viscosity. Note that we have used $z$ to denote the spatial coordinates in the flow examples, hoping that it will not be confused with the Koopman-linear state in \eqref{eq:predictor}.
Similar to \cite{peitz2017koopman}, we assume the forcing $f(z,t)$ is given by
\begin{align}\label{eq:BurgersForcing}
 f(z,t) &= u_1(t) f_1(z)+u_2(t)f_2(z),\\
        &= u_1(t) e^{-\big(15(z-0.25) \big)^2}+u_2(t) e^{-\big(15(z-0.75) \big)^2}
\end{align}
with the  control input $u=(u_1,u_2)\in\mathbb{R}^2$.

The control objective is to follow the reference state
\begin{align}\label{eq:BurgersRef}
 v_{\mr{ref}}(z,0\leq t<2)&= \frac{1}{2}, \nonumber\\
 v_{\mr{ref}}(z,1\leq t<4)&= 1, \nonumber\\
 v_{\mr{ref}}(z,4\leq t<6)& = \frac{1}{2},
\end{align}
starting from the initial condition,
\begin{align}\label{eq:BurgersIC}
 v(z,0)= a e^{-\big(5(z-0.5) \big)^2}+(1-a) \sin(4\pi z).
\end{align}
with $a\in [0,1]$ chosen randomly, and with the input signals constrained as $|u_1|,|u_2|<0.1$.

To construct the Koopman-linear system, we have used 50 two-second long trajectories with $\nu =0.01$.
Each trajectory starts from a random initial condition as in  \eqref{eq:BurgersIC}.
The input control at each time instant is randomly drawn  from the uniform distribution on $(u_1,u_2)\in[-0.1,0.1]^2$. The Burgers equation upwind finite-difference scheme for advection and central difference for diffusion term, with 4th-order Runge-Kutta time stepping performed on 150 spatial grid points with time steps of 0.01 second.

In case of full-state measurements, we use the vector of values of $v$ at the computational grid points, the kinetic energy of $v$, and the constant observable ($\psi(v)=1$). The cost function to be minimized is the kinetic energy ($L^2$-norm) of the state tracking error,
\begin{align}\label{eq:BurgersError}
 e(t) =\int_{0}^{1}|v(z,t)-v_{\mathrm{ref}}(z,t)|^2dz.
\end{align}

For the sparse measurement scenario, we assume that we only have access to the vector of sparse measurements $v^s=\bs{h}(v)$ which consists of values of $v$ at $10$ random grid points. We form the Koopman-linear state vector by including delay embedding of $v^s$ with embedding dimension $n_d=5$, the kinetic energy of instantaneous measurements $\|v^s\|^2$, and the constant observable. That is
\begin{align}
\zeta_c&=[v^s(t_{i-4}),~\ldots,~v^s(t_{i}),~u(t_{i-4}),\ldots,~u(t_{i-1})]^\top,\\
\bs{g}(\zeta_c)&=[\zeta_c^\top,~\|v^s(t_i)\|^2,~1]^\top \in \mathbb{R}^{60}.
\end{align}
 We define the tracking error as
\begin{align}\label{eq:BurgersErrorS}
e(t) =  \frac{1}{m}\|v^s(t)-v^s_{\mathrm{ref}}(t)\|^2,
\end{align}
and the predicted objective function used within the MPC is then
\[
J = \int_0^T e(t)\, dt,
\]
where the prediction horizon is set to $T = 0.1$. After spatio-temporal discretization, this objective readily translates to the form~(\ref{eq:MPCcost}) with $N = 10$.

\begin{figure}[!h] 
	\centerline{\includegraphics[width=1 \textwidth]{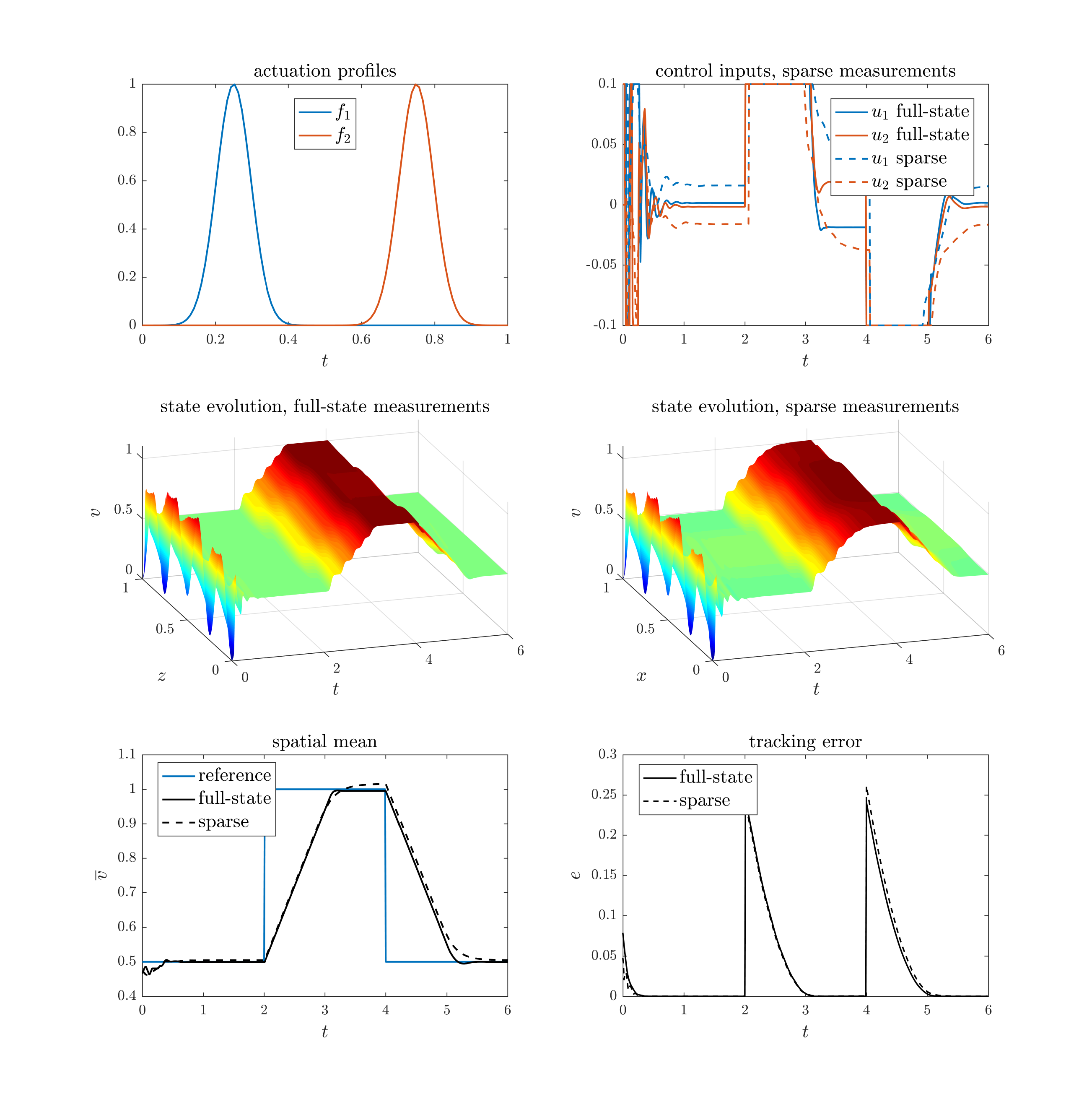}}
	\caption{\footnotesize \textbf{Koopman-MPC for control of Burgers system.}  Input signals, tracking error and state evolution in for closed-loop simulation using Koopman-linear models constructed by full-state measurements (150 observables) and  sparse measurements (10 observables).}
	\label{fig:Burgers1}
\end{figure}

The results of the controlled simulation for both scenarios are depicted in \cref{fig:Burgers1}. In both cases, the state successfully tracks the reference signal; however, the controller built via sparse measurements is slightly delayed compared to the case of full-state measurement which can be attributed to the construction of the Koopman-linear state using delay-embedding. We note that the tracking error during the transient phases is caused by input saturation as documented by the plot of the control input signal.

\begin{figure}[!h] 
	\centerline{\includegraphics[width=1 \textwidth]{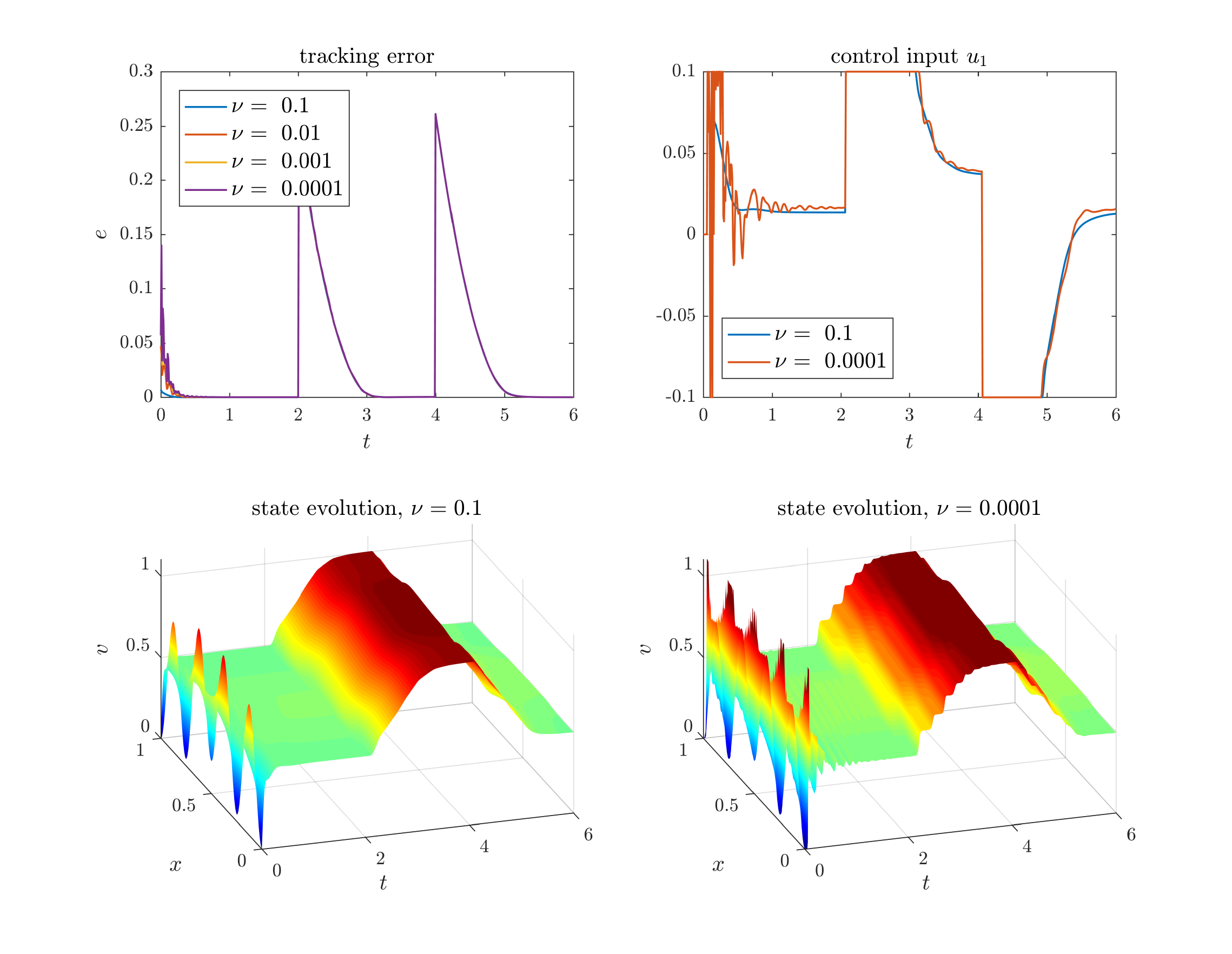}}
	\caption{\footnotesize \textbf{Robustness of Koopman-MPC with respect to parameter $\bs\nu$.} The tracking error and state evolution for controlling the flow at various values of $\nu$ using a controller constructed at $\nu=0.01$ .}
	\label{fig:Burgers2}
\end{figure}

\paragraph{Robustness with respect to the parameter $\bs\nu$.} One question that arises in the context of low-dimensional modeling is wether the models constructed at some parameter value would be robust enough for prediction at other values. In order to test the robustness of Koopman-linear model in case of Burgers, we use the model constructed above using sparse measurements to control the Burgers system at various values of $\nu \in [10^{-4},~0.1]$. The results in \cref{fig:Burgers2} indicate that Koopman-linear model constructed at the parameter regime $\nu=0.01$ is remarkably effective over a wide parameter range, and the control performance is very robust.
As expected, however, the input signal and tracking error in the diffusion-dominated regime (large $\nu$) is less fluctuating, as the diffusion helps the controller to stabilize the state around the spatially-uniform reference state in (\ref{eq:BurgersRef}).

\subsection{2D lid-driven cavity flow}
In the second example, we consider an incompressible viscous flow in a square cavity which is driven by motion of the top lid. The dynamics of the cavity flow is governed by the Navier-Stokes equation, which, in terms of the stream function variable reads
\begin{align} \label{eq:cavityNS}
&\frac{\partial }{\partial t} \nabla^2 \psi+ \frac{\partial \psi}{\partial z_2}\frac{\partial}{\partial z_1}\nabla^2 \psi - \frac{\partial  \psi}{\partial z_1}\frac{\partial}{\partial z_2}\nabla^2 \psi = \frac{1}{Re}\nabla^4 \psi,\quad (z_1,z_2)\in[-1,1]^2,~t\in[0,\infty), \\
&\psi\bigg|_{z_1=\pm1} = \psi\bigg|_{z_2=-1}= 0 \quad \mr{and} \quad \frac{\partial \psi}{\partial z_2}\bigg|_{z_2=1}= f(z_1,t),
\end{align}
where $\psi(z_1,z_2,t)$ is the stream function, $Re$ is the Reynolds number, and $f_1(z_1,t)$ is the velocity of the top lid which acts as the forcing on the system. We assume that we can control the top lid velocity,
\begin{align}\label{eq:CavityForcing}
  f(z_1,t)&= (1+u(t))(1-z_1^2)^2,
\end{align}
with the control input $u\in\mathbb{R}$.

The autonomous cavity flow with $Re\leq 10000$ converges to a steady velocity profile (i.e. fixed point in the state space) which consists of a large central vortex with downstream corner eddies.
At around $Re=10500$, a Hopf bifurcation makes the fixed point unstable and the solutions up to $Re=15000$  converge to a limit cycle. In this regime, the boundary of the central vortex oscillates due to the periodic shedding of vortices from the downstream corners. At higher Reynolds, the flow dynamics grows more complicated and ultimately becomes chaotic at high Reynolds. More details on the dynamics and the numerical scheme used to solve \eqref{eq:cavityNS} can be found in \cite{arbabi2017study}.

We consider two control problems for the lid-driven cavity flow:
\begin{enumerate}[Problem 1), leftmargin=3\parindent]
\item We aim to stabilize the limit cycling flow at $Re=13000$ around the fixed point solution at $Re=10000$. This problem has the trivial solution $u_0=-3/13$, since the effective $Re$ is proportional to the top velocity and $u=u_0$ would set back the flow to the fixed point at $Re=10000$. To avoid the trivial solution, we use the input constraints $-2/13<u<2/13$.
\item Using the same input constraints, we try to stabilize the limit cycling flow at $Re=13000$ around the unstable fixed point solution at the same $Re$. This problem is specially challenging since the linearization around this fixed point has eigenvalues with positive real part that are not controllable. This implies that the nonlinear system is not stabilizable and there is no linear or nonlinear (regular) feedback control that could achieve the full stabilization \cite{sontag2013mathematical}.

\end{enumerate}

The construction of the Koopman-linear system is similar to the Burgers system; we have used 300 two-second long trajectories of the system with control inputs that are randomly drawn from $[-3/13,3/13]$. The initial condition for each trajectory is a random convex combination of the stable fixed point at $Re=10000$, a point on the limit cycle and the unstable fixed point at $Re=13000$. The unstable fixed point is computed via the method proposed in \cite{jordi2014encapsulated}.
 For the full-state observation, we use the values of the stream function on the $50\times 50$ computational grid, the kinetic energy and the constant observable.
For the case of sparse measurement, we use the values of stream functions at $k=2,5,50,100$ random points inside the flow domain, the $l^2$ vector norm of observed stream function values, and the constant observable. According to \cref{sec:DelayEmbed}, the dimension of the state space for the Koopman-linear system built from sparse measurements will be $n = 16,31,256,506$ respectively, which is considerably smaller than the Koopman-linear system with full-state observation ($n=2502$).

Let $\psi_{\mr{ref}}$ denote the stream function at the target fixed point. In the case of the full-state measurements, we define the tracking error to be the kinetic energy of the flow distance from the reference state, i.e.,
\begin{align}\label{eq:MPCerror}
   e(t)=e_k(t):=  \int_\Omega |\mathbf{v}(t)-\mathbf{v}_{\mr{ref}}|^2 d z_1 d z_2,
\end{align}
where $\mathbf{v}=(\p \psi/\p z_2,-\p \psi/\p z_1)$ is the velocity field, and $\Omega$ is the flow domain.

In the case of sparse measurements, let $\psi_s$ be the vector of stream function measurements. Then the tracking error will be the $l^2$-norm of distance from the reference measurements, that is,
\begin{align}
   e(t)=\|\psi_s(t)-\psi_{s,\mr{ref}}\|^2.
\end{align}
The objective function of the MPC is then given by
\[
J = \int_0^T e(t)\,dt,
\]
where the prediction horizon is set to $T = 0.2$. After spatio-temporal discretization, this objective function readily translates to to the form~(\ref{eq:MPCcost}) with $N = 20$.

\begin{figure}[h!]
	\centerline{\includegraphics[width=.55 \textwidth]{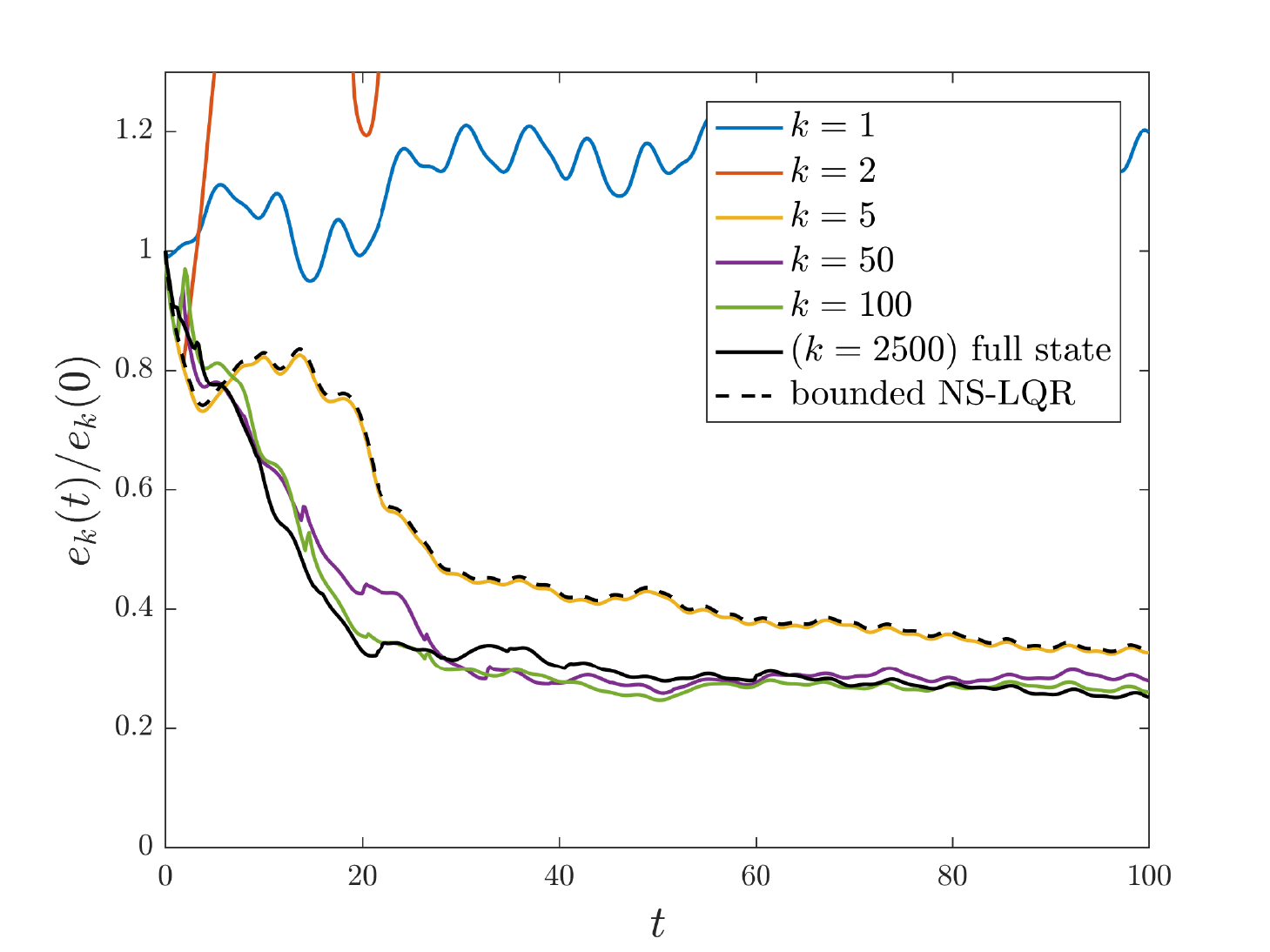}}	\caption{\footnotesize \textbf{Control Performance for stabilization around the steady solution at $\bs{Re=10000}$}. Normalized kinetic energy of flow distance for cavity flow controllers built by Koopman-MPC with various number of measurements ($k$), as well as LQR based on local linearization of Navier-Stokes. }
	\label{fig:CavityError}
\end{figure}

\begin{figure}

\begin{picture}(300,500)
 \put(0,-50){\centerline{\includegraphics[width=1.2\textwidth]{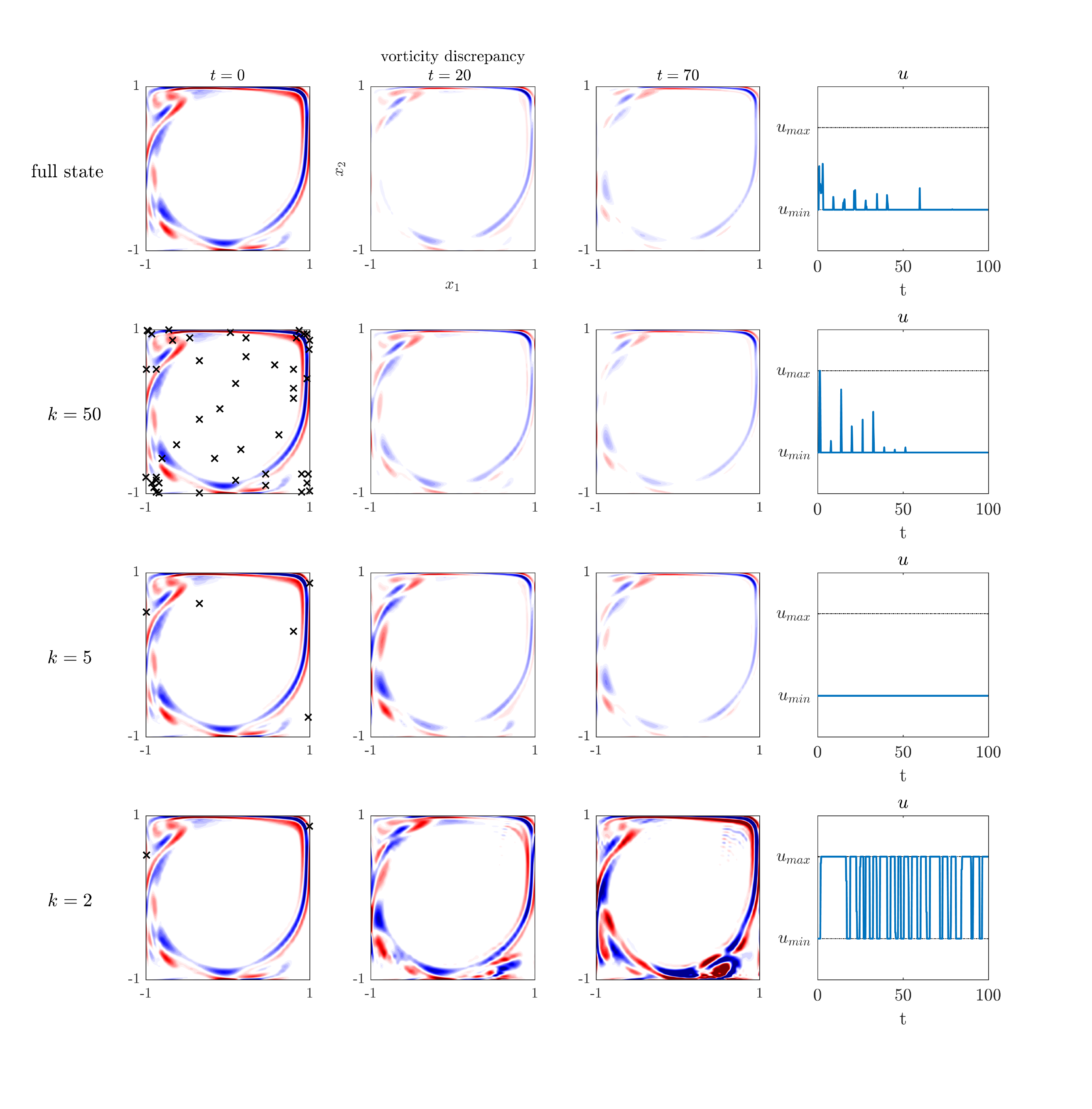}}}
\end{picture}
	\caption{\footnotesize \textbf{Closed-loop control of cavity flow evolution with Koopman-MPC.} Discrepancy in the vorticity of controlled state, and input signal, for full-state and sparse measurements.  The measurement locations are marked via crosses in the leftmost column. The performance of bounded NS-LQR is identical to the Koopman-MPC with $k=5$.}
	\label{fig:CavityVorticity}
\end{figure}

\Cref{fig:CavityError} shows the kinetic energy of the state discrepancy ($e_k$ defined in \eqref{eq:MPCerror})  in applying the Koopman-MPC to problem 1.
All the closed-loop simulations start from the same initial condition on the limit cycle. The Koopman-MPC, except for $k=1,2$, is successful in considerably reducing the flow distance from the desired state over finite time. The control inputs and the flow evolution for some values of $k$ and full-state observation is shown in \cref{fig:CavityVorticity}. In the case of full-state observation, the input signal is mostly saturated at the lower bound which results in a lower effective $Re$ for the flow, and hence getting closer to the fixed point at $Re=10000$. However, the controller  occasionally uses bursts to speed up the stabilization. The effect of these intermittent bursts on the control can be deduced by comparison with the control input with $k=5$ which is saturated at the lower bound at all times.

\Cref{fig:CavityError} also suggests that the control performance of the Koopman-linear systems generally scales with the number of measured observables, i.e., larger number of observables results in better control performance.
This indicates that there is a reasonable tradeoff between the sparsity of measurements and the control performance. Moreover, the  full-state observation offers less than 10 percent improvement over k = 50 in the terminal discrepancy, which indicates that the cavity flow dynamics is approximately low-dimensional and it can be effectively captured using low-dimensional models from data.
We have observed that the choice of measurement location in the flow domain may significantly affect the control performance for very small $k$, e.g. $k=1,2,5$, and results reported in the figures only represent the typical behavior of controllers built on sparse measurements.

A standard technique for flow stabilization is to use linearized Navier-Stokes with linear control strategies. \Cref{fig:CavityError} shows the performance of such technique (labeled bounded NS-LQR) in achieving the control objective of the first problem.
In this method, the Navier-Stokes equation is linearized around the reference fixed point, and an optimal state feedback gain is computed that would minimize the cost function in \eqref{eq:MPCerror} over an infinite-time horizon for the linearized system (see Appendix for detail).  At each time step, the computed optimal input is bounded by the constraints identical to the MPC setting and then applied to the nonlinear system. This method results in an input signal which is saturated at the lower bound and therefore its performance is identical to the case of Koopman-MPC with $k=5$ measurements. This method is successful in substantially reducing the tracking error, but  unlike the Koopman-MPC framework, it is not capable of exploiting the nonlinearities far from the fixed point to speed up the stabilization. Moreover, this method is model-based and its performance is likely to degrade when uncertainties in estimating fluid properties, or input modeling errors are present.

\begin{figure}[h!]
	\centerline{\includegraphics[width=1.1 \textwidth]{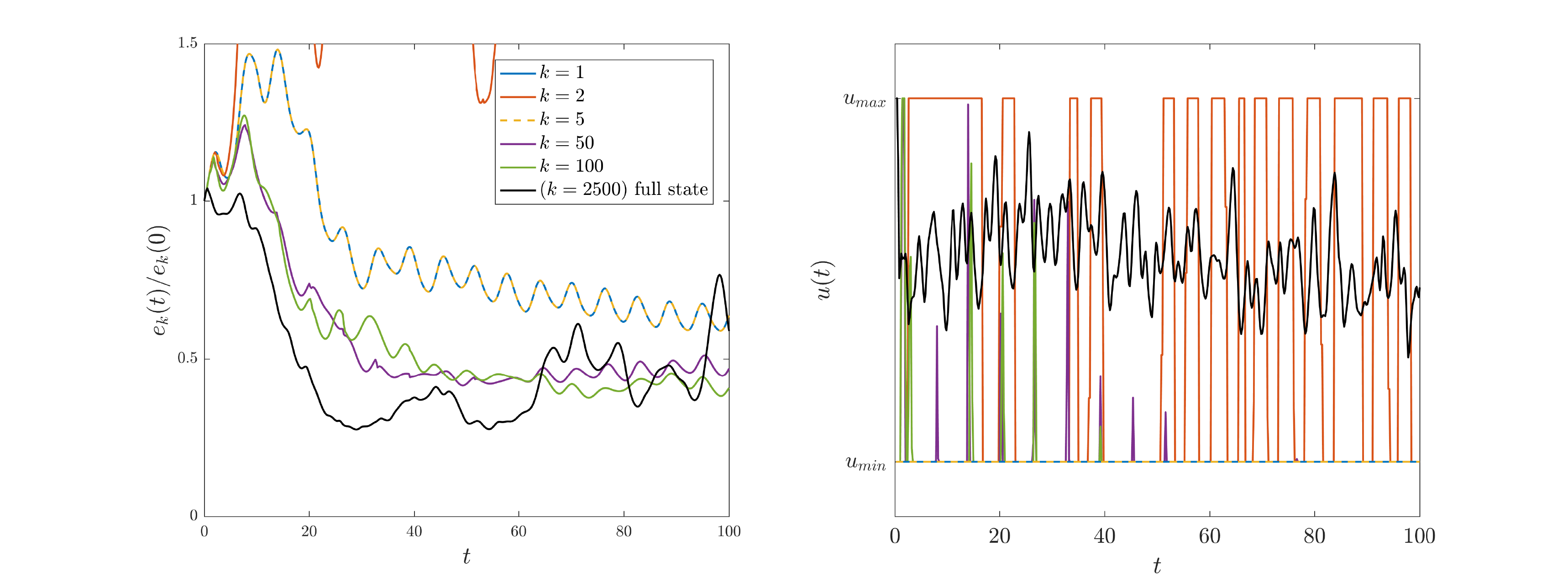}}	\caption{\footnotesize \textbf{Control Performance for stabilization around the steady solution at $\bs{Re=13000}$}: Normalized kinetic energy of state discrepancy and the input signals for cavity flow controllers built by Koopman-MPC with various number of measurements ($k$). The steady solution is not stabilizable and there is no optimal feedback solution for LQR.}
	\label{fig:CavityError2}
\end{figure}

The performance of the Koopman-MPC framework for stabilization around the fixed point at $Re=13000$ (problem 2) is shown in \cref{fig:CavityError2}. Recall that the target fixed point is not stabilizable and no feedback solutions exist that could asymptotically bring the state to the fixed point. Nevertheless, the controllers based on Koopman-linear systems are capable of substantially reducing the tracking error (e.g. down to 40\,\% with $k=100$). The behavior of controllers is similar to the previous problem, i.e., they tend to decrease the effective $Re$ and use occasional bursts to accelerate the stabilization.
An interesting observation is that the controllers built on delay-embedding of measurements perform better than the one with full-state measurements. This shows the effectiveness of using nonlinear observables such as delay-embedded measurements to predict the nonlinear evolution. Note that for this problem, the linearized system around the fixed point is not stabilizable and there is no clear start point for designing feedback control based on linearization techniques commonly used in flow control.

\paragraph{Computation time} \Cref{tab:ControlTime} summarizes the average computational time\footnote{The computations were carried out in MATLAB running on a  3.40 GHz Intel Xeon CPU and 64 GB RAM.} required to evaluate the control input at each time step of the closed-loop operation. We report separately the computation time $t_\mr{embed}$ required to build the state of the Koopman linear system by embedding the available measurements and the time  $t_{\mr{MPC}}$ required to solve the optimization problem~(\ref{eq:MPC}) in the \emph{dense form}\footnote{The conversion of the optimization problem~(\ref{eq:MPC}) to the dense form consist in solving for the state variables $(z_1,\ldots,z_N)$ in terms of the control inputs $(u_0,\ldots,u_{N-1})$ and the initial state $z_0$ using the linear recursion $z^+ = Az+Bu$; the result of this straightforward linear algebra excercise can be found in the appendix of~\cite{korda2018linear}.} using the active set qpOASES solver~\cite{ferreau2014qpoases}. As evident from the table, combination of the Koopman linear representation and convex quadratic programming of the MPC framework leads to computation of the control input in a fraction of a millisecond. Note that in both examples the bulk of the computation time is spent on embedding the sparse measurements to build the Koopman linear state; this step requires data manipulation carried out purely in MATLAB and could be sped up by a tailored implementation (e.g., in~C). Of course, in a real-world implementation of this framework on nonlinear flows, other factors including the time to record and process the physical measurements should also be considered.

\begin{table}[h!]
\centering
\caption{\small \rm Computation time for Koopman MPC of cavity flow and Burgers equations and NS-LQR control for cavity flow. The symbol ``---'' signifies a negligible embedding time in the case of full state measurement.}\label{tab:ControlTime}\vspace{2mm}
\begin{tabular}{ccccc}
\toprule
 & $\#$ of measurements $k$ & embedding dimension $n$ & $t_{\mr{embed}}~[\mr{sec}]$  & $t_{\mr{MPC}}~[\mr{sec}]$\\\midrule
 Burgers &10 & 62 & 4.71$\,\cdot\,10^{-5}$ & 2.28$\,\cdot\,10^{-7}$ \\
&150 & 152 & --- &2.95$\,\cdot\,10^{-7}$ \\\midrule
Cavity &1 & 11 & 1.51$\,\cdot\,10^{-4}$ &8.96$\,\cdot\,10^{-6}$ \\
&2 & 16 & 1.54$\,\cdot\,10^{-4}$ &5.05$\,\cdot\,10^{-5}$ \\
&5 & 31 & 1.52$\,\cdot\,10^{-4}$ &4.19$\,\cdot\,10^{-5}$ \\
&50 & 256 & 1.55$\,\cdot\,10^{-5}$ &3.07$\,\cdot\,10^{-5}$ \\
 &100 & 506 & 1.66$\,\cdot\,10^{-4}$ &7.59$\,\cdot\,10^{-5}$ \\
&2500 & 2502 & --- &4.44$\,\cdot\,10^{-6}$ \\
NS-LQR &2500 & 2501 & --- &$t_{\mr{LQR}}=6.55\,\cdot\,10^{-5}$ \\
\bottomrule
\end{tabular}
\end{table}

\section{Conclusion and outlook}\label{sec:conclusion}
In this work, we discussed the application of the Koopman-linear MPC framework, first proposed in \cite{korda2018linear}, for data-driven control of nonlinear flows.
The key idea is to approximate the Koopman operator from data  to obtain finite-dimensional
linear systems that approximate the nonlinear global evolution of the system and use these systems as the predictor in the model predictive control framework. The combination of Koopman-linear representation of the dynamics and MPC leads to a convex quadratic programming problem that is solved at each time step; this is accomplished using highly efficient and tailored solvers for linear MPC. Moreover, the proposed framework is based solely on data and therefore robust to uncertainties and errors in available models of the nonlinear system. In the problems considered in this work, the Koopman MPC framework showed superior performance compared to feedback strategies based on local linearization and with sub-millisecond computation time.

An important direction for the future work would be to optimize the data collection process to obtain more accurate and efficient Koopman linear models. This requires addressing two problems: first, finding efficient methods for sampling the extended state space of the nonlinear system; in this work we used  random initial condition and random input sequences in the domain of interests to generate data for the EDMD algorithm. The second problem is identifying observables that provide the best finite-dimensional approximation of the Koopman operator in the space of observables. Using machine learning techniques (e.g, \cite{yeung2017learning,takeishi2017learning}) combined with sampling approaches (e.g., \cite{MohrandMezic:2014}) within the Koopman-MPC framework could automatize the choice of observables as well as improve control performance.

\section*{Acknoweldgements}
The authors would like to thank Dr. Sebastian Peitz for a constructive exchange of ideas on the subject.
This research was supported in part by the ARO-MURI
grant  W911NF-17-1-0306, with program managers Dr. Matthew Munson and Dr. Samuel Stanton. The research of M. Korda was partly supported by the Swiss National Science Foundation under grant P2ELP2165166.
\section*{Appendix: Model-based optimal control of cavity flow} \label{app:LQR}
In this section, we describe the design of LQR controller for lid-driven cavity flow based on linearization around the steady solutions. 
Consider the Navier-Stokes equation in \eqref{eq:cavityNS} written as
\begin{equation}
  \frac{\partial}{\partial t}E \psi = f( \psi,u) 
\end{equation}
and let $\psi_0$ be the fixed-point solution corresponding to the input $u_0$, i.e., $f(\psi_0,u_0)=0$.
The linearized Navier-Stokes equations around this equilibrium is given by
\begin{align}
  \frac{\partial}{\partial t}E \tilde \psi = A\tilde{\psi} \label{eq:lin_si_descriptor}
\end{align}
with
\begin{align}
  E:&=\nabla^2 (\cdot),\notag\\
  A:=\frac{1}{Re}\nabla^4(\cdot)-\frac{\partial}{\partial z_2}(\cdot)\frac{\partial}{\partial z_1}\nabla^2 \psi_0 + \frac{\partial  }{\partial z_1}(\cdot) &\frac{\partial}{\partial z_2}\nabla^2 \psi_0 -\frac{\partial \psi_0}{\partial z_2}\frac{\partial}{\partial z_1}\nabla^2 (\cdot) + \frac{\partial  \psi_0}{\partial z_1}\frac{\partial}{\partial z_2}\nabla^2 (\cdot)~. \notag
\end{align}
and $\tilde{\psi}$ is stream function in the linearized equations.
Similar to \eqref{eq:CavityForcing}, the control input to the system is the amplitude of the top velocity lid, which results in the following boundary conditions,
\begin{align}
  \tilde\psi\bigg|_{\partial\Omega}=0,\quad \frac{\partial\tilde\psi}{\partial n} \bigg|_{z_1=\pm1 \text{~or~} z_2=-1}=0,\quad \text{and}\quad \frac{\partial\tilde\psi}{\partial z_2} \bigg|_{z_2=+1}= u(t)(1-z_1^2)^2, \label{eq:lin_bc}
\end{align}
with $\tilde{u}$ as the deviation from the base input $u_0$.

In order to transform the boundary control problem into the standard linear time-invariant (LTI) format, we introduce the extension function
  \begin{align}
  H(z_1,z_2)=\frac{1}{4}(1-z_1^2)^2(1+z_2)^2(z_2-1),
\end{align}
and use the change of variables
\begin{align}
 \eta (z_1,z_2) = \tilde \psi(z_1,z_2) - H(z_1,z_2)\tilde{u}(t),
\end{align}
The linear system in the new variable reads
\begin{align}
    \frac{\partial}{\partial t} E \eta = A\eta +  AH \tilde{u} - EH\frac{d\tilde{u}}{dt},
\end{align}
 with homogeneous boundary condition,
 \begin{align}
  \eta \bigg|_{\partial\Omega}=\frac{\partial\eta}{\partial n}\bigg|_{\partial\Omega}=0. \label{eq:etaBC}
\end{align}
The above is in fact an LTI \emph{descriptor} system which can be written as
  \begin{align*}
  \frac{\partial}{\partial t}
     \begin{bmatrix}
    E && 0 \\ 0 && 1
  \end{bmatrix}
  \begin{bmatrix}
    \eta \\ \tilde{u}
  \end{bmatrix}
  =
    \begin{bmatrix}
    A && AH \\ 0 && 0
  \end{bmatrix}
  \begin{bmatrix}
    \eta \\ \tilde{u}
  \end{bmatrix}
  +
  \begin{bmatrix}
    -EH \\ 1
  \end{bmatrix}
  \frac{d\tilde{u}}{d t}.
\end{align*}
or in a more compact form,
\begin{align}
  \mathbf{E}\dot{\x}=\mathbf{A}\x + \mathbf{B}\dot{\tilde{u}},\label{eq:linsys1D}
\end{align}
where $\x=[\eta,\tilde{u}]^\top$ is the embedded state.
We are interested in finding the optimal input $d\tilde{u}/dt$ (and $\tilde{u}$)  for the above system that minimizes the cost function,
  \begin{align}\label{eq:LQRcost}
  J(\x,\dot{u}) &=\int_0^\infty\bigg[ \int_\Omega |\mathbf{v}|^2 dA + \frac{}{}\alpha_1\tilde{u}^2+\alpha_2(\dot{\tilde{u}})^2\bigg] dt
\end{align}
where $\mathbf{v}=(\p\tilde{\psi}/\p z_2,-\p \tilde{\psi}/\p z_1)$ is the velocity field of the linearized system.


We have used the Chebyshev collocation scheme \cite{trefethen2000spectral} to spatially discretize the linear system in \eqref{eq:linsys1D} and the cost function in \eqref{eq:LQRcost} . We have chosen $\alpha_1=\alpha_2=10^{-6}$ to minimally penalize the input and avoid infinitely large solutions.
If the linear system is stabilizable (for example in the case of steady solution at $Re=10000$), solving the continuous-time algebraic Riccati equation (ARE) (see e.g. \cite{arnold1984generalized} for descriptor formulation of ARE), would give the optimal feedback gain $k=[k_\eta~~k_u]$. The LQR optimal input $\tilde{u}$ is then computed by time stepping the following ordinary differential equation
  \begin{align*}
\dot{\tilde u}= -k_uu-k_\eta\eta= -k_u\tilde u-k_\eta(\tilde\psi-H\tilde u),
\end{align*}
and the input $u=u_0+\tilde{u}$ is applied to the nonlinear system. If the fixed point is not linearly stabilizable, such as the steady solution at $Re=13000$, then ARE does not have a solution and there is no stabilizing input.

\bibliography{FlowControl.bbl}
\bibliographystyle{unsrt}
\end{document}